\documentclass{JHEP3}
\usepackage{amsfonts}
\usepackage{amsmath}
\usepackage{amssymb}
\usepackage{latexsym}
\usepackage{graphicx}
\usepackage{cite}
 
\newcommand\no{\nonumber\\{}}

\newcommand\eqnb{\begin{eqnarray}}
\newcommand\eqne{\end{eqnarray}}

\newcommand\void[1]{}   
\newcommand{\Tr}{{\mathrm{Tr}}}

\usepackage{latexsym}
\numberwithin{equation}{section}

\def\cN{\mathcal{N}}

\def\cR{\mathcal{R}}

\newcommand{\be}{\begin{equation}}
\newcommand{\ee}{\end{equation}}
\newcommand{\bea}{\begin{eqnarray}}
\newcommand{\eea}{\end{eqnarray}}

\def\calD{{\mathcal D}}

\preprint{ DAMTP-2010-26}

\title{5 loops in 24/5 dimensions}

\author{Jonas Bj\"ornsson and Michael B. Green\\
 Department of Applied Mathematics and
Theoretical Physics\\
Wilberforce Road, Cambridge CB3 0WA, UK\\
\email{\tt J.Bjornsson@damtp.cam.ac.uk\\ M.B.Green@damtp.cam.ac.uk}}

\abstract{A first quantised approach to loop amplitudes based on the  pure spinor particle is applied  to the systematics of four-particle amplitudes in maximally supersymmetric field theories.  Counting of fermionic zero modes  allows the identification of momentum factors multiplying  $\cR^4$ in the case of supergravity (and  $F^4$ in the Yang--Mills case) thereby making manifest their ultraviolet properties as a function of dimension, $D$.
For $L=2,3,4$ loops the leading supergravity divergence is in $D=4+6/L$ dimensions and proportional to $\partial^{2L}\, \cR^4$, in line with earlier field theory calculations.  However, at five loops there is a radical change in the systematics, suggesting the presence of a contribution with an explicit $L=5$ logarithmic ultraviolet divergence when $D=24/5$  that is proportional to $\partial^8\cR^4$.  We further argue that  $\partial^8\cR^4$ should receive contributions from all loops, which would imply that $\cN=8$ supergravity (with $D=4$)   is not protected by supersymmetry from a seven-loop divergence.  

}

\keywords{pure spinor; supersymmetry, supergravity}

\begin{document}

%\maketitle

%%%%%%%%%%%%%%%%%%%%%%%%%%%%%%%%%%%%%%%%%%%%%%%%%%%%%%%%%%%%%%%%%%%%%%%%%%%%

\section{Introduction}

Maximally supersymmetric Yang--Mills and supergravity have been the focus of a great deal of attention over the years.  In particular, four-dimensional $\cN=4$ Yang--Mills theory  is special since it is a perturbatively finite quantum field theory, which also arises as the low energy limit of open string theory and, in the large-$N$ limit (with gauge group $SU(N)$), is related to closed-string string theory via the gauge/gravity correspondence.   However,  $\cN=8$ supergravity is non-renormalisable in four dimensions, and its scattering amplitudes probably do not have a consistent  perturbative formulation  because of ultraviolet divergences,  although the order at which such divergences appear  --  indeed, whether they appear at all -- is still a controversial issue.  Explicit evaluation of  the four-graviton scattering amplitude has demonstrated the absence of such divergences for $L=1,2,3,4$ (where $L$ is the loop number), but there is a strong suspicion that although low order terms are protected by maximal supersymmetry, divergences will appear at higher orders.   

The degree of divergence of loop contributions to scattering amplitudes is correlated, by simple dimensional analysis, with the  power of the external momenta that can be extracted into the prefactor multiplying the loop integral.   This is a useful way of describing the ``critical'' dimension, $D=D_c^{(L)}$, which  is the lowest dimension in which the amplitude is ultraviolet divergent at $L$ loops.  For example, the one-loop ($L=1$) contribution to the  four-graviton amplitude in maximal supergravity has a factor of $\cR^4$, which contains eight  powers of momentum,   multiplying a box diagram of  $\varphi^3$ scalar field theory.   Since this contains eight inverse powers of momentum\footnote{Here $\cR$ is the linearised curvature, which is quadratic in momentum, and the four curvatures are contracted in a well-known manner that may be  determined by supersymmetry.}, the amplitude is less divergent in the ultraviolet than its naive degree of divergence.   In a sufficiently high dimension, $D> D_c^{(1)}$,  there is an ultraviolet divergence that  behaves as $\Lambda^{D-8}$, where $\Lambda$ is a momentum cutoff.  This  becomes a logarithmic divergence when $D=D_c ^{(1)}= 8$, which would be seen as a pole in $\epsilon$ in dimensional regularisation in $D=8+2\epsilon$ dimensions.  At higher loops there are more powers of external momenta.  It follows from a simple dimensional argument that the $L$-loop amplitude in $D> D_c^{(L)}$ dimensions has a leading low-energy behaviour  that  can be expressed  as\footnote{In this paper we will not obtain the precise normalisations, so we will ignore many constant factors.}  
\be
A^{(L)} = \sigma_2^{p_L}\, \sigma_3^{q_L}\, \cR^4\, \Lambda^{L(D-2) -6 -2\beta_L}\,,
\label{loopamp}
\ee
where $\beta_L \equiv 2p_L+3q_L$ 
\be
\sigma_n = s^n + t^n + u^n
\label{signdef}
\ee
  (and $s$, $t$, $u$ are  Mandelstam invariants, which are quadratic in the momenta).
 The expression  \eqref{loopamp} is the most general Bose symmetric scalar invariant that can be made out of  $2\beta_L$ powers of the four external momenta.  It follows that the logarithmic ultraviolet divergence appears when 
 \be
 D= D^{(L)}_c = 2 +  {6+ 2\beta_L\over L} \,.
 \label{firstuv}
 \ee
The kinematic factors in the low-energy loop amplitudes translate into  terms in the effective action of the form $\partial^{2\beta_L} \, \cR^4$, where we  have suppressed the precise pattern of contractions of derivatives and of the four Riemann tensors.  
  Analogous arguments apply to the momentum factors in maximally supersymmetric Yang--Mills theory, although the effects of colour ordering distinguish the single-trace contribution, $\Tr F^4$, and the double-trace contribution, $(\Tr F^2)^2$, to four-particle scattering.

 Various arguments suggest that, at least for small values of $L>1$, the power of momenta is given by 
 \be 
 \beta_L = L\,,
 \label{linear}
 \ee
 so that 
 \be
 D_c^{(L)} = 4 + {6\over L}\,.
 \label{criticald}
 \ee
  This  is the result of  explicit evaluation of  four-graviton scattering amplitudes in maximal supergravity at up to $L=4$ loops  in~(\cite{Bern:1998ug,Bern:2007hh,Bern:2009kd} ($L=1$ being special \cite{Green:1982sw}).   
 These explicit calculations demonstrate that the loop amplitudes can be expressed as sums of terms with external powers of momentum multiplying scalar loop diagrams, that superficially look like a certain subset of $\varphi^3$ scalar field theory diagrams.  For $L=1$ the scalar diagram is simply the box diagram while for $L=2$ there is a crossing symmetric combination of planar and nonplanar double-box diagrams.  However, for $L= 3,4$ important extra numerator factors of loop momenta are inserted is a specific pattern.  The presence of these internal momenta implies the diagram is more divergent than a corresponding $\varphi^3$ diagram would be (after all, $\varphi^3$ is ultraviolet finite for $D<6$, which is surely not the case for supergravity).

The pattern of ultraviolet divergences beyond four loops depends on whether an extra factor of $s\sim \partial^2$ continues to  arise in the prefactor for each extra loop beyond $L=4$, which would mean that the relation $\beta_L=L$ continues to hold\footnote{Note that if $\beta_L=L$ for all $L$ then $D_c^{(L)} = 4+ O(1/L)$, so the theory would be free of ultraviolet divergences to all orders~\cite{Green:2006gt}.}.  For example, if this does continue there should be a five-loop logarithmic divergence proportional to $\partial^{10}\, \cR^4$ in $D=26/5$ dimensions, whereas if the five-loop amplitude is proportional to $\partial^8\, \cR^4$ the logarithmic divergence would arise in $D=24/5$ dimensions.

The established  pattern of derivatives acting on $\cR^4$ up to $L=4$ is in accord with a variety of arguments, some of which are based on use of string/M-theory,  showing that terms of the form $\partial^{2L}\, \cR^4$ with $L<4$, $\cR^4$, $\partial^4\, \cR^4$, $\partial^6\, \cR^4$,  are protected from getting contributions beyond $1, 2$ and $3$ loops respectively  (with analogous statements for $F^4$)~\cite{Green:1997tv,Green:1998by,Berkovits:2004px,Green:2005ba,Berkovits:2006vc}.  Underlying these results is the intuition that these interactions are protected by supersymmetry by virtue of being $1/2$-BPS, $1/4$-BPS and $1/8$-BPS operators.  As stressed in \cite{Green:2010sp}, a corollary is that $\partial^8\, \cR^4$ is not  BPS protected and therefore likely to get perturbative corrections at all loops.  

A powerful framework for addressing the  constraints imposed by maximal supersymmetry is the pure spinor formalism \cite{Berkovits:2000fe,Berkovits:2000wm}, which was formulated as a method for calculating string theory amplitudes in a manifestly supersymmetric manner.  In this formalism the classical world-sheet fields, which are space-time superfields,  are supplemented by fermionic and bosonic  pure spinor ``ghost'' fields.  Integration over fermionic zero modes leads to explicit momentum factors in the low energy limit of string loop amplitudes.  However, there are practical difficulties in using this formalism for general amplitudes beyond two loops due to singularities at small values of the bosonic pure spinor field $\lambda$, which are postponed to higher loops for four-point functions~\cite{Berkovits:2006vc}. The onset of these small-$\lambda$ singularities is connected with the order in the derivative expansion at which interactions are protected from renormalisation by string loop corrections. In \cite{Berkovits:2006vc} it was  suggested that interactions of the form $\partial^{2k}\, \cR^4$ in the closed-string four-graviton amplitude do not receive loop corrections for $L>k$ when $1<k\le 5$.   However, this analysis overlooked subtle zero-mode effects, pointed out  in \cite{Berkovits:2009aw}, which indicated that this nonrenormalisation holds only for $k=2,3$.  In the open string case this was used to explain the fact that the double trace $\cN=4$ Yang--Mills interaction,  $\partial^2\, (\Tr F^2)^2$,   is an ``F-term'' that is protected from renormalisation for $L\ge 3$ (whereas  $\partial^4\, (\Tr F^2)^2$ is a ``D-term'' that gets contributions for all $L\ge 1$).  In the closed-string case this is in accord with the intuition that $\partial^8\, \cR^4$ is not protected and, consequently, there  is likely to be a seven-loop divergence in  $\cN=8$ supergravity with $D=4$ (instead of a nine-loop divergence that was argued for in~\cite{Green:2006yu}).

In this paper we will study properties of maximally supersymmetric field theory making use of a formulation of   pure spinor quantum mechanics presented in more detail in paper \cite{jonas}.  This extends and generalises~ \cite{Berkovits:2001rb,Anguelova:2004pg} to multiloop amplitudes.
The aim is to clarify the powers of momenta multiplying $F^4$ and $\cR^4$ in four-particle amplitudes  rather than evaluating the precise expressions for the amplitudes. As in the case of the superstring, the most efficient way of describing maximal supersymmetry is to use the ten-dimensional language.  For that reason we will often refer to  supersymmetric Yang--Mills theories as the $\cN=1$ case and maximal supergravity as the $\cN=2$ case. We will focus mainly on the supergravity amplitude, with a few remarks about the Yang--Mills case. The notation and methods are  based on analogies with the corresponding pure spinor string theory, although several new issues arise.  Although we will be directly concerned with the field theory analysis of multi-loop diagrams, many of the arguments are closely based on the structure of pure spinor string theory.  Consequently, we will be able to pinpoint the subtleties of the analysis of the non-renormalisation properties much more precisely than the sketchy arguments presented in  \cite{Berkovits:2009aw}.

The expressions for $L$-loop amplitudes in the pure spinor particle formalism will be reviewed in 
section~\ref{review}.
This will include a description of the zero modes of the world-line fields that play a key r\^ole in determining the powers of momenta in the prefactors multiplying the dynamical part of the amplitude.  The detailed structure of the four-particle loop amplitudes  will be described in section~\ref{multiloop}.   The aim here is to determine the powers of momenta in the prefactors and properties of the scalar field theory diagrams multiplying them.  This provides information concerning the leading ultraviolet properties of the amplitude.  For $L\le 4$ we reproduce the structure of the amplitudes that have already been evaluated using other methods~\cite
{Green:1982sw,Bern:1998ug,Bern:2007hh,Bern:2009kd}. In the pure spinor formalism the powers of momenta in the prefactor are determined through a supersymmetric mode counting argument.   For $L=5$ there is a qualitative change in the combinatorics of the zero modes.  We will show that in this case the mode counting procedure leads to a contribution 
to $\partial^8\, \cR^4$, thereby breaking the pattern of terms of the form $\partial^{2L}\, \cR^4$ that holds for $L=2,3,4$.
Furthermore, we will give a sketchy argument that $\partial^8\,\cR^4$ also gets contributions from loops with $L >5$.  Since we have not evaluated the precise values of the coefficients the possibility of terms vanishing or  cancellations between different contributions to the amplitude cannot be ruled out.  Nevertheless, such cancellations would have to be quite different in nature from those  that arise for $L=3,4$ in  \cite
{Bern:2007hh,Bern:2009kd},  and would not be consequences of supersymmetry in any conventional sense.
We will end with a short discussion of these results in section~\ref{discussion}.

\section{Loop amplitudes and pure spinor quantum mechanics}
\label{review}

The action for the pure spinor particle with ten-dimensional $\cN=1$ supersymmetry   in the minimal formalism was introduced in \cite{Berkovits:2001rb}. This may be extended to the ``non-minimal'' formalism (introduced in the string context in \cite{Berkovits:2005bt})  and generalised to $\cN=2$ supergravity  by mimicking the transition from the open string to the closed string by doubling all the fields in \cite{Berkovits:2001rb} apart from $X$ and $P$. The particle action in this $\cN=2$  non-minimal formalism is~\cite{jonas}
\eqnb
S &=& \int_{F_L} d\tau \left( \dot{X}P + \dot{\theta} p + \dot{\lambda} w + \bar{w}\dot{\bar{\lambda}}-s\dot{r}- \hat{p}\dot{\hat{\theta}} + \hat{w}\dot{\hat{\lambda}} + \dot{\hat{\bar{\lambda}}}\hat{\bar{w}} + \dot{\hat{r}}\hat{s}-\frac{P^2}{2}\right),
\label{spinact}
\eqne
where the integral is over the network of world-lines that forms a particular ``skeleton diagram'', denoted ${F_L}$ (a skeleton is a Feynman diagram with the external legs removed). Equation \eqref{spinact} is the analogue of a gauge-fixed particle action in which the moduli are encoded in the lengths of the world-lines joining vertices in the network.  The world-line fields in this action consist of the classical superspace phase-space  coordinates together with a set of pure spinor ghosts.  The classical coordinates are the bosons,  $X^m$, $P_m$ (where $m=0,\dots,9$) and fermions  $\theta^\alpha$, $p_\alpha$ ($\alpha=1,\dots,16$).  The  non-minimal  $\cN=1$   ``pure spinor ghost''  coordinates and their momenta consist of the bosonic fields $\lambda^\alpha$,  $w_\alpha$, $\bar \lambda_\alpha$,  $\bar w^\alpha$ and the fermionic fields $r_\alpha$, $s^\alpha$.   All the fields apart from $X^m$ and $P_m$ are doubled by the hatted fields, leading to $\cN=2$ supersymmetry.\footnote{In the following we will concentrate on the unhatted world-line fields since the properties of the hatted fields are identical.}  We will use the IIA notation in which the hatted field $\hat\theta_\alpha$ has a lower index.  The conjugate  momenta, $P_m$, $w_\alpha$, $\bar w^\alpha$ and $s^\alpha$ are world-line vectors, which have $L$ zero modes on an $L$-loop skeleton, while the other fields are world-line scalars, which have a single zero mode. Note in particular that the loop momenta in a skeleton are the zero modes of $P_m$.

The Hamiltonian that follows from the action is $H = P^2/2$, so the equations of motion for all fields other than $X^{m}$ have solutions that are locally constant.  The world-line fields $\lambda^{\alpha}$, $\bar{\lambda}_\alpha$ and $r_\alpha$ are pure spinors that satisfy the constraints
\be
\lambda\gamma_{m}\lambda = 0\,, \qquad 
\bar{\lambda}\gamma_{m}\bar{\lambda} = 0\,,\qquad
\bar{\lambda}\gamma_{m}r = 0,
\ee
which imply that $\lambda$, $\bar{\lambda}$ and $r$ each have eleven degrees of freedom.  
Therefore,  only the following  linear combinations of $w_\alpha$, $\bar w^\alpha$ and $s^\alpha$ are invariant under the gauge transformations implied by the pure spinor constraints,
\eqnb
J &=& \lambda w\,,\qquad \bar{J} = \bar{w}\bar{\lambda} - sr \,,\no
N_{mn} &=& \frac{1}{2} w\gamma_{mn}\lambda\,, \qquad
\bar{N}_{mn} = \frac{1}{2}\bar{w}\gamma_{mn}\bar{\lambda}- \frac{1}{2}s\gamma_{mn}r \,,\no
S &=&s\bar{\lambda} \,, \qquad S_{mn} = \frac{1}{2}s\gamma_{mn}\bar{\lambda}\,.
\label{gaugecomb}
\eqne

The BRST charge for the $\cN=2$ theory is the sum of two $\cN=1$ supercharges,
\be
Q_{tot} = Q + \hat Q\,,
\label{qtot}
\ee
where
\eqnb
Q = \lambda d + \bar{w}r\,, \qquad \hat Q=\hat{d}\hat{\lambda} + \hat{r}\hat{\bar{w}}\,,
\eqne
and where $d_\alpha=p_{\alpha} + P_m (\gamma^m\, \theta)_\alpha/2$, with a similar formula for $\hat d^\alpha$.
A composite $b$-ghost, $b_{tot}$, can be constructed by mimicking the procedure of \cite{Berkovits:2005bt} in the superstring.  This is the sum of two pieces,  $b_{tot}=b +\hat{b}$ and is constrained to satisfy
\eqnb
[Q_{tot}, \, (b + \hat{b})]=\left[Q,b\right] + [\hat{Q},\hat{b}] =  H\,,\qquad [Q_{tot}, \, (b- \hat{b})]=\left[Q,b\right] - [\hat{Q},\hat{b}] = 0\,,
\eqne 
 which are the zero mode parts of the closed string relations.  The components  $b$ and $\hat b$ also satisfy $b^2= \hat{b}^2=0$. 
The solution for $b$ follows closely the string expression and is given by\footnote{It should be noted that, just as in the case of the pure spinor string, the condition $b^2=0$  with $b$ defined by  \eqref{bghostN1} has only been partially checked.}
\eqnb
b &=& \frac{1}{2}\left(\frac{G^\alpha\bar{\lambda}_{\alpha}}{\left(\lambda\bar{\lambda}\right)} + \frac{\bar{\lambda}_{\alpha}r_\beta H^{[\alpha\beta]}}{\left(\lambda\bar{\lambda}\right)^2} + \frac{\bar{\lambda}_{\alpha}r_\beta r_\gamma K^{[\alpha\beta\gamma]}}{\left(\lambda\bar{\lambda}\right)^3} + \frac{\bar{\lambda}_{\alpha}r_\beta r_\gamma r_\delta L^{[\alpha\beta\gamma\delta]}}{\left(\lambda\bar{\lambda}\right)^4}\right),
\label{bghostN1}
\eqne
where $G$, $H$, $K$ and $L$ are the zero modes of the expressions that arise for the pure spinor string in equation (3.16) in \cite{Berkovits:2005bt},
\eqnb
G^\alpha
	&=&
		-\frac{1}{2}P^m\left(\gamma_m d\right)^\alpha\,, \qquad
H^{[\alpha\beta]}
	=
		-\frac{1}{384}\gamma^{\alpha\beta}_{mnp}\left[\left(d\gamma^{mnp}d\right) - 24 N^{mn}P^p\right] \no
K^{[\alpha\beta\gamma]}
	&=&
		\frac{1}{192}\gamma_{mnp}^{[\alpha\beta}\left(\gamma^md\right)^{\gamma]}N^{np} \,,\qquad
L^{[\alpha\beta\gamma\delta]}
	=
		\frac{1}{12244}\gamma_{mnp}^{[\alpha\beta}{{\gamma^m}_{qr}}^{\gamma\delta]}N^{np} N^{qr}\,.
\label{bfields}
\eqne
The $\hat{b}$-ghost is given by the same expression using hatted instead of unhatted world-line fields. 

It will  be important to understand the r\^ole of the terms in \eqref{bghostN1} in the mode counting arguments in the next section.  The pattern we will find is the following.  For loop-amplitudes with $L=1$ and $L=2$ only the second term, with coefficient $H$ (containing a term quadratic in $d$)  contributes to the full amplitude for both the Yang--Mills and supergravity cases. For $L\ge 3$ the situation is more complicated and we will only discuss the leading contribution in the low energy limit (which is also the term with the leading ultraviolet divergence).  For $L=3$ and $L=4$  only the first two terms in \eqref{bghostN1}  contribute to the leading low energy behaviour of the supergravity amplitude.  In the Yang--Mills case the third term also plays a r\^ole for $L\ge 3$. However, we will see that for $L\ge 5$ the mode counting changes in such a way that all terms in \eqref{bghostN1} are relevant. The only  condition is that the total number of $r$'s is more then eleven in the product of the $b$-ghosts.  

As with the string, we need to consider unintegrated and integrated vertex operators, $U$ and $V$, respectively.   These satisfy
\be
[Q,U(X,\theta,\hat{\theta})] = [\hat{Q},U(X,\theta,\hat{\theta})]=0\,,
\label{brstunintvert}
\ee
and
\be
[Q,V(X,\theta,\hat{\theta})] = [H,K]\, \quad[\hat{Q},V(X,\theta,\hat{\theta})] = -[H,\hat{K}]\,.
\label{brstintvert}
\ee
These equations are solved by
\be
U\left(X,\theta,\hat{\theta}\right) = \lambda^{\alpha}{A_{\alpha}}^{\beta}\left(X,\theta,\hat{\theta}\right)\hat{\lambda}_{\beta}\,,
\label{unintvert}
\ee
\eqnb
V\left(X,\theta,\hat{\theta}\right) 
	&=&
		P^m G_{mn} P^n + d_{\alpha}{W^{\alpha}}_{\beta}\hat{d}^{\beta}- d_{\alpha}{\hat{E}^{\alpha}}_{m}P^m - P^m{E}_{m \alpha}\hat{d}^{\alpha} 
	\no
	&-&
		 \frac{1}{2}N^{mn}\hat{C}_{mn;\alpha}\hat{d}^{\alpha} - \frac{1}{2} d_{\alpha} {C^{\alpha}}_{mn}\hat{N}^{mn}  
	\no
	&+& 
		\frac{1}{2}N^{mn}\hat{\Omega}_{mn;p}P^p + \frac{1}{2}P^m \Omega_{m;np}\hat{N}^{np}+ \frac{1}{4} N^{mn}S_{mn;pq}\hat{N}^{pq}\,,
		\label{intvert}
\eqne
where the space-time fields ${A_\alpha}^{\beta}$, $G_{mn}$, ${W^\alpha}_{\beta}$, $\dots$ are functions of the superspace coordinates $(X^m, \theta^\alpha, \hat \theta_{\alpha})$ and satisfy the linearised field equations of type IIA supergravity in ten dimensions. The vertex operators was determined in \cite{Berkovits:2001ue,Grassi:2004ih}. The metric superfield $G_{mn}(X,\theta,\hat \theta)$ contains the curvature in the term $\theta^2\, \hat\theta^2\, \cR$, while the bispinor superfield, ${W^\alpha}_{\beta}$, contains the curvature in the term $\theta\, \hat \theta\, \cR$, since $W \sim D \hat D \, G$, where the covariant derivative is given by $D_{\alpha} = \partial_{\alpha} + (\gamma^m\theta)_{\alpha}\partial_{m}/2$ (with a similar expression for $\hat D$). Furthermore ${A_\alpha}^{\beta}$ contains the curvature in the term $\theta^3\, \hat\theta^3\, \cR$. The fields ${\hat{E}^{\alpha}}_{m} $ and ${E}_{m{\alpha}}$ contain the curvature in the terms $\theta\hat{\theta}^2\cR$ and $\theta^2\hat{\theta}\cR$, respectively.  Higher order terms in the $\theta$, $\hat \theta$ expansions of these superfields contain derivatives of the curvature in the usual way.
An important observation is that there are momentum factors $P^m$ in parts of the vertex operators which, when they come in pairs, can produce contact terms in the amplitudes.  

 In determining the properties of the amplitudes in the next section we will need to understand the r\^oles of the different components of the vertices, \eqref{unintvert} and \eqref{intvert}.  We will see that for $L=1,2$ only the $d\, W \,\hat d$ part of the integrated vertex,  $V$, contributes to the amplitude (the unintegrated vertex, $U$, also enters the $L=1$ case).  For $L=3, 4$ the first four terms in $V$ contribute.  For $L\ge 5$ the leading term in the low energy limit (the term that is most divergent in the ultraviolet) only gets contributions from the $P^m\, G_{mn}\, P^m$ part of $V$.

\begin{figure}[t]
\centerline{\rotatebox{0}{\scalebox{0.2}{\includegraphics*{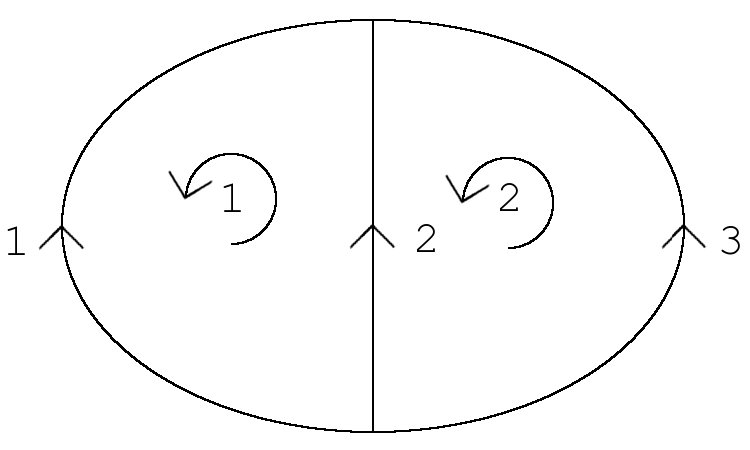}}}}
\caption{The unique two-loop skeleton diagram. The amplitude is obtained by attaching vertex operators to points on the lines, which are integrated around the diagram.   The circular arrows denote the different $b_I$-cycles. The propagators in the skeleton are numbered from 1 to 3 and the arrows on each line indicate the direction of increasing proper time along the line.}
\label{Figure:twoloop}
\end{figure}

As a preliminary to describing the pure spinor amplitudes, we will review the first quantised  description of  scalar $\varphi^3$ loop diagrams, which is based on \cite{Dai:2006vj} and is analogous to the structure of bosonic string loop amplitudes. Any $L$-loop skeleton diagram has $(3L-3)$ internal lines, which are parameterised by choosing an origin for each line and which have lengths $T_i$ ($i=1,\dots, 3L-3$).     A basis of ``$b_I$-cycles'' is associated with a choice of $L$ inequivalent internal counter-clockwise loops. An example at two loops is shown in figure \ref{Figure:twoloop}. One can use the $b_I$-cycles to parametrize the non-trivial one-forms of the skeleton, $\omega_I\equiv{a_{I}}^{i} d\tau_i$ where ${a_{I}}^{i}$ is equal to $\pm 1$ if $d\tau_i$ is in the same/opposite direction to the $b_I$-cycle. As the zero modes of the world-sheet vectors are expanded using the non-trivial one-forms, one finds $L$ zero modes for each world-line vector field. This plays an  important r\^ole for the pure spinor particle amplitudes as we will see later on.  The period matrix,  $\Omega_{IJ}$, is a symmetric matrix defined by $\Omega_{IJ}\equiv \oint_{b_I}\omega_{J} = \int_{{F_L}}\omega_{I}\omega_{J}/d\tau$, which appears to have $(L^2+L)/2$ components. However, at most $3L-3$  of these are linearly independent (for $L>1$).

  A loop amplitude contributing to the scattering of $N$ scalar particles is constructed by attaching   vertex operators $V_0(k_r,\tau_r) \equiv e^{ik_rX(\tau_r)}$  for  particles with momentum $k_r$  at points $\tau_r$  ($r=1,\dots, N$)  on the skeleton and performing the functional integral over $X^m$ in the standard manner.   In principle, we should include an integral over the $(b,c)$  gauge-fixing ghosts, but this simply gives an overall factor that we will ignore. The amplitude is given by
\eqnb
I(\{T_i\},\{\tau_r\},  \{k_r\}) &=&  \left< V_0(k_1, \tau_1)\dots
V_0(k_N,\tau_N) \right>_{F_L} \no 
& =&
		\int d^L\ell e^{-\int d\tau \frac{1}{2}\left(\ell^I\omega_{I}/d\tau\right)^2}e^{-\sum_{r<s}^N k_rk_sG\left(\tau_r,\tau_s\right)}\delta\left(\sum_{r=1}^N k_{r}\right)\no
	&=&
		\frac{\left(2\pi\right)^{DL/2}}{\Delta^{D/2}}e^{-\sum_{r<s}^N k_rk_sG\left(\tau_r,\tau_s\right)}\delta\left(\sum_{r=1}^N k_{r}\right),
\label{scalarparticleloop}
\eqne
where
\be
 \left< V_0(k_1, \tau_1)\dots V_0(k_N,\tau_N) \right>_{F_L} = \int \mathcal{D}X\, \prod_{r=1}^N e^{ik_rX(\tau_r)}e^{-S_B} \,,
\label{matdef}
\ee
and $S_B$ is the bosonic particle action, $D$ is the dimension of space-time and $\Delta \equiv \det\left(\Omega_{IJ}\right)$.  
In \eqref{scalarparticleloop} the loop momenta $\ell^I_m$ are the $L$ zero modes of $P_m$ for the skeleton diagram, i.e., $\left. P_m\, d\tau\, \right|_{zero}  = \sum_I  \ell^I_m \, \omega_I$.  

The  Green function $G\left(\tau_r,\tau_s\right)$ can be determined, using an electric circuit analogy  in the form \cite{Dai:2006vj}
\eqnb
G\left(\tau_i,\tau_j\right)
	&=&
		-\frac{1}{2}\left|\int_cds\right|+\frac{1}{2}\int_c\omega_I \left(\Omega^{-1}\right)^{IJ}\int_c\omega_J,
\label{Greensfunc}
\eqne
where $c$ in any non-intersecting path between $\tau_i$ and $\tau_j$.
The bosonic amplitude is obtained by integrating the $\tau_r$'s over the skeleton and then integration over the moduli, $T_i$,
\be
A(\{k_r\}) = \int_0^\infty dT_1 \dots dT_{3L-3}    \, \int_{{F_L}} \prod_{r=1}^N d\tau_r\, I(\{T_i\},\{\tau_r\},  \{k_r\})\,,
\label{bosamp}
\ee
where $\int_{F_L}$ indicates that the positions of the vertex operators are to be integrated independently  around all the lines of the skeleton.   In the low energy limit,  $k_r \to 0$,  this expression is ultraviolet divergent at sufficiently high $D$.  Introducing a momentum cutoff $\Lambda$ (or a small-$T_i$ cutoff)  the amplitude is proportional to $\Lambda^{L(D-6)+6- 2N}$, which defines the degree of divergence of $\varphi^3$ in $D$ dimensions.

The multi-loop amplitudes for maximal supergravity are constructed in  \cite{jonas} by analogy with the structure of the pure spinor string theory loop amplitudes of \cite{Berkovits:2005bt}.   The tree-level ($L=0$)  amplitudes and one-loop ($L=1$)  amplitudes  are special since they have at least one unintegrated vertex operator, while for $L>1$ only the integrated vertices appear. For $L>1$, BRST invariance requires the insertion of $(3L-3)$ $b$-ghost insertions multiplying  ``Beltrami differentials'', which reduce to a factor of $\int_0^{T_i} d\tau b(\tau)/T_i$ for each line of the skeleton. As a result, the generalisation of the bosonic functional integral \eqref{scalarparticleloop} is (for $L>1$)\footnote{We are here denoting the vertex $V(X,\Theta)$ for a plane wave of momentum $k_r$ by $V(k_r,\tau_r)$.}
\eqnb
&&\hspace{-0.5cm} K(\{T_i\},\{\tau_r\},  \{k_r\}) \no
&& \hspace{-0.2cm} =
	\int \calD X DP\int \mathcal{D}\Phi\mathcal{D}\hat{\Phi}\left(\mathcal{N}\mathcal{\hat{N}} \prod_{i=1}^{3(L-1)}\left(\int_0^{T_i}\frac{d\tau}{T_i}b\int_0^{T_i}\frac{d\tau}{T_i}\hat{b}\right)V(k_1, \tau_1)\ldots V(k_4,\tau_4)\, e^{-S}\right)\,.
	\no\label{N=2loop.def}
\eqne
Here $\mathcal{D}\Phi\mathcal{D}\hat{\Phi}$ denotes the measure for the functional  integral over fermionic and pure spinor world-line fields.   This generalises the bosonic functional integral \eqref{scalarparticleloop} to include integration over the fermionic and pure spinor variables.  
The amplitude is again obtained by integrating over the positions of the vertices and the values of the moduli,
\be
A(s,t,u) = \int_0^\infty dT_1 \dots dT_{3L-3}    \, \int_{{F_L}} \prod_{r=1}^4 d\tau_r\,  K(\{T_i\},\{\tau_r\},  \{k_r\})\,.
\label{supamp}
\ee

As in the pure spinor string, in the absence of a regulator the integral  apparently diverges at the boundary where $\lambda$, $\bar{\lambda}$, $\hat{\lambda}$ or $\hat{\bar{\lambda}}$ are large, but also has a vanishing factor since there are not enough fermionic zero modes to saturate the Grassmann integrals.  This $0/0$ ambiguity is dealt with by introducing the regulator $\mathcal{N}\hat{\mathcal{N}}$,  where
\eqnb
\mathcal{N}
	&=&
		e^{q_1 \int_{{F_L}} d\tau \left[Q,\chi^1\right]+ q_2 \int_{{F_L}} d\tau\left[Q,\chi^2\right]}
	\no
	&=&
		e^{-q_1\int_{{F_L}}d\tau\left(\lambda\bar{\lambda}-\theta r\right)
	-
		q_2\int_{{F_L}}d\tau\left(N_{mn}\bar{N}^{mn}+J\bar{J}-\frac{1}{2}\left(d\gamma_{mn}\lambda\right)S^{mn}-\lambda d S\right)}\,
\label{regulator}
\eqne
is the exponential of a BRST-exact quantity and $q_1,q_2$ are arbitrary positive constants.  In writing the second line  the choices
$\chi^1 =  \theta\bar{\lambda}$ and $\chi^2= \left(N^{mn}S_{mn} + JS\right)$ have been made. The hatted regulator, $\hat \cN$, is defined the by the same expression with hatted variables replacing the unhatted ones.    

The expression \eqref{N=2loop.def} is not quite BRST invariant due to boundary contributions arising from collisions of single-particle vertex operators attached to the same line of the skeleton.  This lack of BRST invariance is cured by including multi-particle contact vertices that couple two or more external particles at the same point on a line of the skeleton.  The form of these vertices is determined explicitly in \cite{jonas}.  Such contact terms are needed for both Yang--Mills and supergravity. For the case with four external particles, only four-point vertices arise and has a non-zero contribution for amplitudes with $L\geq 3$ (see  \cite{jonas} for more details). In the case of Yang-Mills, each contact term reduces the number of external momenta of the prefactor in the low energy limit by one whereas for supergravity, it does not effect the low energy limit.  Furthermore, as in the case of the pure spinor string, a problem arises with a singularity at small $\lambda$ at a sufficiently large number of loops, $L$, which needs to be regulated as in~\cite{Berkovits:2006vi,Aisaka:2009yp} (a somewhat different regulator was proposed in~\cite{Grassi:2009fe}).

\medskip
{\bf Counting fermionic zero modes}
\smallskip

The measure of the fermionic and pure spinor zero modes is of the form (see  equations (4.4), (4.6) and (4.18) of  \cite{Berkovits:2005bt} for an exact breakdown of the measure)
\be
D\Phi_0 =  \bar{\lambda}^{8L+3}\left(\lambda\bar{\lambda}\right)^{-(8L+3)} \, d^{10L}N\, d^{10L}\bar{N}\, d^L J\, d^L \bar{J}\, d^{11L}S\,  d^{16L}d\,  d^{11}\bar{\lambda} \, d^{11}\lambda\,  d^{11}r\,  d^{16}\theta\,,
\label{measure}
\ee
with a similar expression for $\mathcal{D}\hat\Phi_0$. 
It is useful to write this formally as 
\eqnb
\mathcal{D}\Phi_0 \sim d^{11L}\, \bar{w}\,  d^{11L}w\,  d^{11L}s\,  d^{16L}d\,  d^{11}\bar{\lambda} \, d^{11}\lambda\,  d^{11}r\,  d^{16}\theta\,.
\label{zeromeasure}
\eqne  
Although this ignores crucial factors arising from the pure spinor constraints  it is useful as a mnemonic for counting the independent modes. 

The integration over fermionic modes proceeds as follows.  The only source of the $11L$ $s$ fields needed to saturate the integral  is the regulator, $\cN$,   where each $s$  is multiplied by a $d$.   We also see from the measure  that at least $5L$ further $d$'s are needed,  5 for each $b_I$-cycle in the skeleton diagram, in order to saturate the $d$ zero modes.    These can only come from the $b$-ghost insertions, each providing at most two $d$'s,  and from the vertices, which can each provide one $d$.  Here and in the following we explicitly describe the insertion of single-particle external vertices, making comments on the effect of contact vertices that  couple pairs of external states, where appropriate. Powers of $r$ arising from the $b$-ghost insertions are traded for covariant derivatives, $D_\alpha$,  with respect to $\theta^\alpha$, due to the factor $e^{q_1\theta r}$ in the regulator \eqref{regulator}.    Similar considerations apply to the counting of the hatted modes.
  This counting plays a crucial r\^ole in constraining the contributions of the scalar loop diagram and  is the key to determining  the number of derivatives acting on $F^4$ or $\cR^4$, and hence in determining the ultraviolet dependence of the diagrams.
  
  As we are particularly interested in zero modes, it is enlightening to consider the case in which  the composite $b$-ghost insertion only contains zero modes of the constituent fields. In this case we can express $b$ in terms of a second-rank tensor, $\left. b\right|_{zero} =  b^{IJ}\,\omega_I \omega_J/(d\tau)^2$, giving
\eqnb
\int_0^{T_i}\frac{d\tau}{T_i}\, \left . b\, \right|_{zero}
	=
		b^{IJ}\frac{1}{T_i}\int_0^{T_i}\omega_{I}\omega_{J}/d\tau
	=
		b^{IJ}\frac{\partial\Omega_{IJ}}{\partial T_i}\,.
\eqne
Note that $b^{IJ}$   has the same number of independent components as $\partial\Omega_{IJ}/ \partial T_i$, which is at most $3L-3$.    This dependence on the period matrix  shows that the amplitude depends crucially on the topology of the skeleton.  The parts of $b$ containing nonzero modes of its component fields will be  discussed later.
 
 An essential issue in analysing any diagram is whether the product of $b$ zero mode insertions,
\eqnb
\prod_{i=1}^{3L-3} \int_0^{T_i}\frac{d\tau}{T_i}\left .b\right|_{zero} 
		= b^{I_1J_1} \dots b^{I_{3L-3} J_{3L-3}}\,   \frac{\partial\Omega_{I_1J1}}{\partial T_1}\dots  \frac{\partial\Omega_{I_{3L-3} J_{3L-3}}}{\partial T_{3L-3}} \,,
		\label{prodb}
\eqne
 can contribute with the maximum number of $d$ zero modes, via $H^{[\alpha\beta]}$ in the second term in \eqref{bghostN1}.   Schematically, for this term we may write  $b_H^{IJ} \sim d_\gamma^I\, d_\delta^J$ (where $\gamma,\delta=1,\dots, 5$ label the five components of the spinor indices that have not been matched with the indices on the $s$'s).  When $b^{IJ} \to b^{IJ}_H$  the expression~\ref{prodb} is  only nonzero  if
 \be
 \sum_{i=1}^{3L-3} c_i\, \frac{\partial \Omega_{IJ}}{\partial T_i}\ne 0\,,
 \label{linearomega}
 \ee
 for  any nonzero constants $c_i$ because $(b_H^{IJ})^2=0$ for all $(I,J)$.  For diagrams in which \eqref{linearomega} is not true there must be contributions from nonzero modes of $b$, in which case the $b$ insertions provide fewer $d$ zero modes than  when \eqref{prodb} is satisfied and so more $d$ zero modes must come from the vertex insertions.  This provides extra constraints on the positions at which vertices may be attached to the skeleton.
 The situation changes when $L\ge 5$ where it is necessary to regulate a small-$\lambda$ divergence~\cite{Berkovits:2006vi}.

After integration over the non-minimal fields the amplitudes can be reduced, in the low energy limit,  to matrix elements of the form 
\eqnb
\left.\left<  \lambda^3\hat\lambda^3\, {\cal O} \right>\right|_{\theta^5\hat{\theta}^{5}},
\label{minmatrix}
\eqne
which were introduced in \cite{Berkovits:2000fe,Berkovits:2000wm}. This denotes the integration over the minimal world-line scalar fields $\lambda$ and $\hat{\lambda}$ and the fermionic fields $\theta$ and $\hat{\theta}$ and this picks out a  particular $\theta^5\hat{\theta}^5$ component of the operator ${\cal O}$.

\section{Multi-loop systematics}
\label{multiloop}

We will now describe the systematics of the integration over zero modes for the $L$-loop four-particle amplitudes.  These  determine the momentum factors  as well as  the structure of the scalar loop integrals.  As summarised in the introduction, this information determines the leading ultraviolet properties of the amplitude, which  can be compared to the known maximal supergravity results for $L=1$ \cite{Green:1982sw}, $L=2$ \cite{Bern:1998ug}, $L=3$ \cite{Bern:2007hh} and $L=4$ \cite{Bern:2009kd}.   The zero mode analysis allows us to pinpoint features of the scalar loop diagrams, such as the distribution of vertices attached to the internal lines and the origin of momentum numerator factors, which are otherwise obscure. Comments will also be made about the structure of the maximally supersymmetric Yang--Mills amplitudes.
 
 An important feature of the analysis will involve understanding the way in which the saturation of fermionic zero modes restricts the manner in which different components of the vertices and the $b$-ghost contribute.  
 We saw earlier that these have to provide $5L$ factors of $d$.  As a result,
 for $L=1,2$ the only part of $V$ that can contribute is the term $d\, W \, \hat d$ in \eqref{intvert}, while for $L>2$ the other terms in $V$ enter.  These terms each contain zero, one or two  factors of $P^m$.  Similarly, for $L=1,2$, in order to saturate the fermionic modes only the second term in the expression for $b$ in \eqref{bghostN1} contributes to the amplitude.  However, for $L>2$ other terms in $b$ that again contain factors of $P^m$ contribute.   The lesson is that for $L>2$ new kinds of contributions arise in which factors of $P^m$  enter into the bosonic part of the functional integral.   Therefore, the  bosonic factor in the functional integral, the  expectation value in \eqref{scalarparticleloop},  needs to be generalised to allow for the inclusion of insertions of up to two powers of $P^m$ at each vertex position, together with possible factors of $P^{n_r}(\rho_r)$ arising from the $b$-ghost insertions at $\rho_r$,
\be
 \left< P^{n_1}({\rho_1}) \dots P^{n_s}(\rho_s)\, V(k_1,\tau_1) \dots V(k_4,\tau_4) \right>_{F_L}\,,
 \label{modP}
 \ee
 where $V(k_r, \tau_r)$  is equal to either  $e^{ik_r\cdot X(\tau_r)}$,  $P^m_r(\tau_r) \, e^{ik_r\cdot X(\tau_r)}$ or $P^{m_r}(\tau_r)\, P^{n_r}(\tau_r)\, e^{ik_r \cdot X(\tau_r)}$.
  For the supergravity case there is an even number, $2q$, of $P^m$ insertions.  The expression  \eqref{modP} is therefore a $2q$-index tensor which contracts with  a tensor arising from integration over the fermion and pure spinor variables.  The full result also requires a sum over all possible insertions.
  
   The bosonic integral \eqref{modP}  can be evaluated making use of the contraction between $P^m$'s,
 \eqnb
\left<P^{m}(\tau)P^{n}(\tau')\right> &=& - \delta^{mn}\omega_I(\tau)\left(\Omega^{-1}\right)^{IJ}\omega_J (\tau')/(d\tau d\tau')\,.
\label{contractp}
\eqne
Although the result is complicated in general,  its low energy limit simplifies since all the factors of $P^m$ contract with each other in pairs, giving, in dimensions $D> D_c^{(L)}$, an expression of the form
\be
 \left<P^{m_1}(\tau_1) \dots P^{m_{2q}}(\tau_{2q})  \right>_{F_L}\sim \Lambda^{L(D-6) +6 -2N+2q}\,,
\label{presu}
\ee
where all the indices reduce to products of Kronecker deltas, and $N=4$ in the absence of contact terms.
 Therefore, the pattern of $P^m$ insertions is a key part in determining the leading ultraviolet divergences.  
 In the Yang--Mills case the number of $P^m$ insertions need not be even. When the number is odd,  the extra $P^m$ gets traded for an external momentum in the leading low energy contribution. Furthermore, the presence of contact terms in Yang--Mills effectively decreases the value of $N$.
   
We will see that the nature of the mode counting changes drastically for $L=5$, in a manner that was mentioned earlier.  At this point there are sufficient $b$-ghost insertions  to cause a small-$\lambda$ singularity that has to be regularised, although we only make minimal use of the rather complicated regulator of ~\cite{Berkovits:2006vi,Aisaka:2009yp}.  The result is that specific skeleton diagrams with specific distributions of zero modes are responsible for the leading $\partial^8\, \cR^4$ low energy behaviour of the amplitude when $L\ge 5$.

\medskip
{\bf (i)  One loop. }
\smallskip
 In the $L=1$  case five $d$ zero modes are required to come from the vertices and the single $b$ insertion (recalling that $11$ $d$'s having been used to multiply the 11 $s$'s).  
The $L=1$ skeleton is simply the trace of the logarithm of the  propagator, to which  the four vertex operators are attached, which introduces four propagators,   Three of these vertices are integrated around the circle, while the fourth is an unintegrated vertex.    The unintegrated vertex provides no $d$'s, so the three integrated vertices must each provide one $d$ and so they contribute via the $W$ term in \eqref{intvert}. The $b$ insertion provides the two remaining $d$'s through the second term in \eqref{bghostN1}, as well as a factor of $r$, which gives a factor of $D$ after integration.   Similar comments apply to the $\hat b$ insertion and the $\hat d$ modes\footnote{Generally in the following the counting of  hatted modes follows that of the unhatted ones.}. The  amplitude reduces to a scalar box diagram multiplying the factor coming from integration over the fermionic and pure spinor variables and  at low energy is proportional to
\be
A^{(1)} \sim  \left.\left<\lambda^3\hat{\lambda}^3D\hat{D} A W^3\right>\right|_{\theta^5\hat{\theta}^{5}} \, \Lambda^{D-8} \sim D^8 \hat D^8 \, G^4 |_{\theta=\hat \theta=0}\, \Lambda^{D-8} \sim  \cR^4 \, \Lambda^{D-8}\,,
\label{rfouramp}
\ee
in the supergravity case (and $F^4\,  \Lambda^{D-8}$ in the Yang--Mills case), 
where  the scalar product is  defined in \eqref{minmatrix}.  It is clear that~\eqref{rfouramp}  can clearly be written as an integral over half the superspace, which is consistent with $\cR^4$  being a $1/2$-BPS interaction.  The amplitude has a logarithmic ultraviolet divergence when $D=8$. This example  was analysed in detail in \cite{Mafra:2005jh} for the pure spinor string. 

\medskip
{\bf (ii)  Two loops}
\smallskip

 There is  also a unique skeleton for the $L=2$ amplitude.   Furthermore, the  ten missing $d$ zero modes must arise from the four integrated vertices and three $b$ insertions.   This  means that each vertex contributes through the $W$ term  and each $b$ once more gives two powers of $d$.   The requirement that the vertices contribute four $d$ zero modes, with two attached to each loop,  leads to the constraint that a maximum of two vertices can be attached to each line of the skeleton, which coincides with the distribution found in \cite{Bern:1998ug}. The supergravity amplitude in this case is given by  the two-loop skeleton with scalar vertices attached multiplied by  the fermion and  pure spinor factor, which gives the low energy behaviour
\be
A^{(2)} \sim \left.\left<\lambda^3\hat{\lambda}^3D^3\hat{D}^3 W^4\right>\right|_{\theta^5\hat{\theta}^{5}}\, \Lambda^{2(D-7)} \sim D^{12}\, \hat D^{12}\, G^4 \, \Lambda^{2(D-7)}\sim 
\sigma_2\,\cR^4  \, \Lambda^{2(D-7)}\,,
\label{d4r4}
\ee
and  $\partial^2 \Tr F^4\, \Lambda^{2(D-7)}$ and $\partial^2 (\Tr F^2)^2\, \Lambda^{2(D-7)}$ for the single-trace and double-trace terms in the Yang--Mills amplitude.
 The two-loop pure spinor superstring amplitude was determined in detail in \cite{Berkovits:2005ng}. The form of \eqref{d4r4} illustrates that $\partial^4 \, \cR^4$ arises as an integral over $3/4$ of  the superspace and is therefore a $1/4$-BPS interaction, which has a logarithmic ultraviolet divergence when $D=7$.

\medskip
{\bf (iii)  Three loops}
\smallskip

\begin{figure}[t]
\begin{center}
\begin{tabular}{cc}
{\rotatebox{0}{\scalebox{0.2}{\includegraphics*{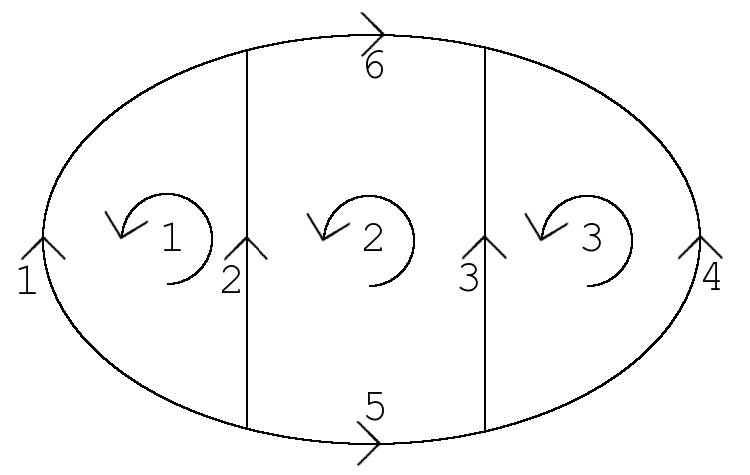}}}}
& 
{\rotatebox{0}{\scalebox{0.2}{\includegraphics*{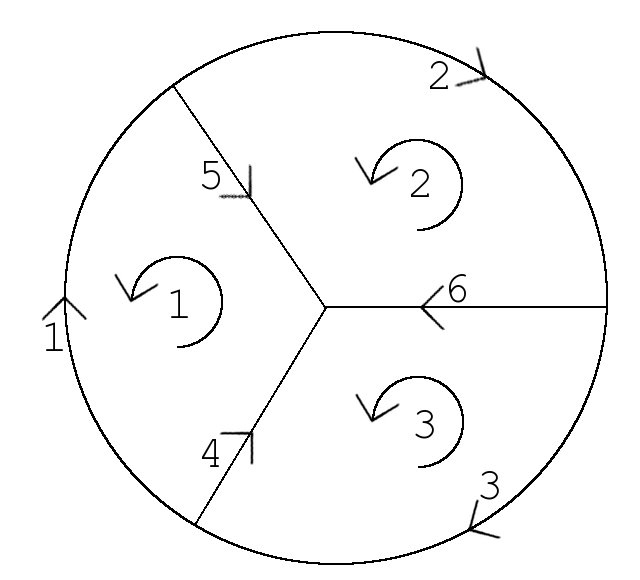}}}}
\\ (a) & (b)
\end{tabular}
\caption{(a) The three-loop ladder skeleton. (b) The ``Mercedes'' skeleton.}
\label{Figure:threeloop}
\end{center}
\end{figure}
When $L=3$  there are two skeleton diagrams -- the ladder and Mercedes diagrams shown in  figure~\ref{Figure:threeloop}.  In this case there are 15 missing $d$ zero modes, which must come from the vertices and the 6 $b$ mode insertions.  The same holds for the hatted modes.  The two skeletons give rise to very different  mode counting, which follows from the structure of the period matrix. The period matrix for the ladder in figure~\ref{Figure:threeloop}(a) is
\eqnb
\Omega_{IJ}
	&=&
		\left(
		\begin{array}{ccc}
			T_1 + T_2 & -T_2 & 0 \\
			-T_2 & T_2 + T_3 + T_5 + T_6 & -T_3 \\
			0 & -T_3 & T_3 + T_4
		\end{array}
		\right),
		\label{ladderthree}
\eqne
while for the Mercedes diagram in figure~\ref{Figure:threeloop}(b) it is 
\eqnb
\Omega_{IJ}
	&=&
		\left(
		\begin{array}{ccc}
		T_1 + T_4 + T_5 & -T_5 & -T_4 \\
		-T_5 & T_2 + T_5 + T_6 & -T_6 \\
		-T_4 & -T_6 & T_3 + T_4 + T_6
		\end{array}
		\right)\,.
\eqne
The ladder period matrix  has one less nonzero independent element than the Mercedes matrix, which is important for the distribution of the $b$-ghost components.   For this diagram  $\Omega_{IJ}$ in \eqref{ladderthree}  does not satisfy the condition \eqref{linearomega}.  In this case 
 the $b$-ghost insertions cannot all contribute with the maximal number of $d$ zero modes, via $b^{IJ}_H\sim d^I_\delta d^J_\gamma$ in the second term in \eqref{bghostN1}, because  the $b$ insertions on lines $5$ and $6$  in figure~\ref{Figure:threeloop}(a) would  both have to equal $b_H^{22}$  (see the discussion after \eqref{prodb}).
The $b$ insertion on one of these  can only contribute at most a single $d$ zero mode.   One such contribution
 arises by inserting the first term in $b$  \eqref{bghostN1} (the $G^\alpha \bar\lambda_{\alpha}$ term) in line 5.  For this term $b^{22} \sim d^2_\gamma\, \ell^2_{(5)}$, where $\ell^2_{(5)}\equiv\ell^2$ is the momentum in line 5.  Adding the analogous contribution from the same insertion on line 6  gives a total insertion proportional to 
 $ d^2_\gamma\, k_m$, where $k_m$ is total external momentum flowing through lines 5 and 6. 

It follows that the maximum number of $d$ zero modes that can be obtained from $b$ insertions  is  3 for loops 1 and 3, and 5 for loop 2,  giving a total of 11 $d$ zero modes. Therefore, each single-particle vertex has to contribute the maximal number of $d$ zero modes (a total of 4),  which means that they must all be $W$ vertices, which do not have factors of $P^m$. Furthermore,  one pair of vertices must be attached to the first loop  and the other pair to the third loop.    Simple dimensional counting determines the leading low momentum dependance of the amplitude to be proportional to $\partial^8\cR^4 \,\Lambda^{3(D-6)-2} $. Other contributions in which one $b$ insertion has a single $d$ zero mode are also possible and has the same consequences.  
  
\begin{figure}[t]
\begin{center}
{\rotatebox{0}{\scalebox{0.2}{\includegraphics*{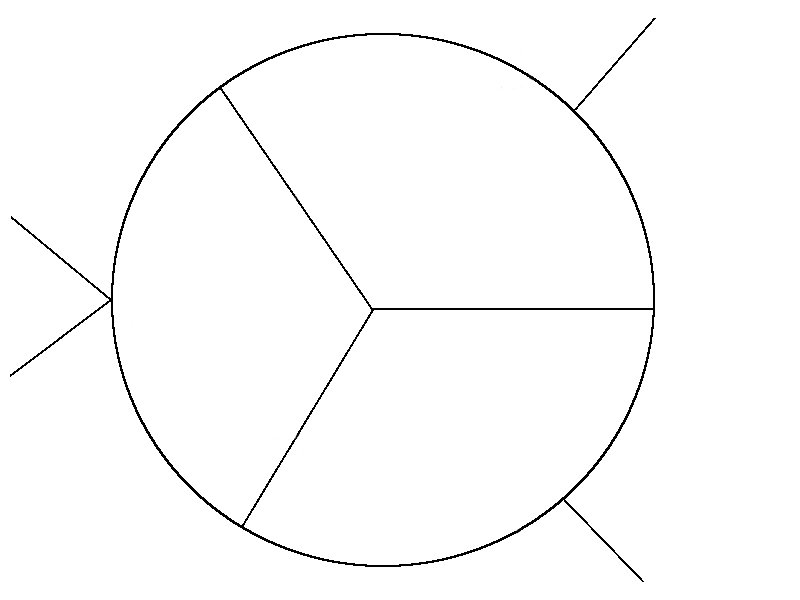}}}}
\caption{A three-loop diagram with one contact term that arises in maximal supergravity and in maximal Yang--Mills.  While its contribution makes no qualitative change to the leading behaviour of the supergravity amplitude, in the Yang--Mills case its presence is responsible for the leading behaviour,  $\partial^2\, \Tr F^4$.}
\label{Figure:threecontact}
\end{center}
\end{figure}

For the Mercedes diagram the $b$-ghost insertions can contribute the maximal number of $d$ zero modes since the number of nonzero independent elements of the period matrix is 6 so that  \eqref{linearomega} is satisfied.  With this configuration the 6 $b$ insertions contribute 12 $d$'s,  4 for each loop.  Only three more $d$'s are needed from the vertex insertions. As a consequence,  one vertex need not contribute a $d$ and another vertex (or the same vertex) need not contribute a $\hat d$. The net result is that the leading low energy term in the amplitude is one in which the  vertices contribute a factor of  $P^mP^n$ via the first term in \eqref{unintvert}.  This reduces at low energy to a factor of $P^2$ in the scalar field theory factor in the functional integral, as discussed in \eqref{presu}.  Furthermore, at least one vertex must be attached to each loop. In other words,  the leading part of the amplitude at low momenta is proportional to the Mercedes diagram with insertions of scalar vertices and one power of $P^2$.  This multiplies the fermionic and pure spinor integrals.  There are a variety of ways in which the 3 $d$'s and 3 $\hat d$'s  can be distributed among the four single-particle vertices, which all lead to a leading contribution of the same order.   For example, one of these arises from a configuration with single $E$ and  $\hat E$  factors from  the $V$ vertices,  giving the low-energy dependence 
\be
A^{(3)} \sim\left.\left<\lambda^3\hat{\lambda}^3D^6\hat{D}^6 E \hat{E} W^2\right>\right|_{\theta^5\hat{\theta}^{5}} \, \Lambda^{3(D-6)}\sim
D^{14}\, \hat D^{14}\, G^4\, \Lambda^{3(D-6)}
\sim \sigma_3\, \cR^4\, \Lambda^{3(D-6)}\,,
\label{d6r4}
\ee
which is equivalent to an integral over $7/8$ of superspace and is therefore a $1/8$-BPS interaction.  Apart from the above there are several different contributions that give the same behaviour in the low energy limit, which arise by choosing one $b$ insertion to have the first term in \eqref{bghostN1},  instead of the second.  This term has a factor of $P^m$ but only has one  $d$, so in this case all the vertices have to contribute one $d$.  There are similar possibilities for the hatted insertions.  Other configurations contribute expressions of the same form.  These include, in particular, the contribution shown in figure~\ref{Figure:threecontact} that involves a contact interaction attaching two external states to one of the lines of the skeleton. As mentioned earlier, such contact terms are important for BRST invariance when $L\ge 3$, but in the supergravity case they do not alter the degree of divergence.
Note that the presence of  a $P^2$ insertion in each Mercedes diagram corresponds to the internal momentum numerator factor in the analysis of \cite{Bern:2007hh} and is responsible for the $\Lambda$-dependence in (\ref{d6r4}), which implies a logarithmic ultraviolet divergence in the critical dimension $D_c^{(3)} =6$.  
 
   In the  the maximally supersymmetric Yang--Mills  case the vertex operators can contribute a single factor of $P^m$ in the Mercedes diagram. It is straightforward to see that this solitary momentum factor becomes a factor of an external momentum after the functional integration, which gives  an amplitude proportional to $\partial^4F^4$. However, this is not the complete story since  the contact interactions required by  BRST  determine the leading behaviour in the Yang--Mills case.    One insertion of a contact vertex  is needed in the Mercedes diagram  (see figure \ref{Figure:threecontact}), which reduces the dimension of the single-trace part of the amplitude to  $\partial^2\Tr F^4\, \Lambda^{3(D-6)}$. However, this contact term does not contribute to the double-trace part, which remains proportional to $\partial^4(\Tr F^2)^2 \Lambda^{3D-20}$,  as observed in \cite{Dixon:2009tk, Bern:2010tq}.  This coincides with the conclusions in \cite{Berkovits:2009aw}.

\medskip
{\bf (iv)  Four loops}
\smallskip

\begin{figure}[t]
\begin{center}
\begin{tabular}{ccc}
{\rotatebox{0}{\scalebox{0.13}{\includegraphics*{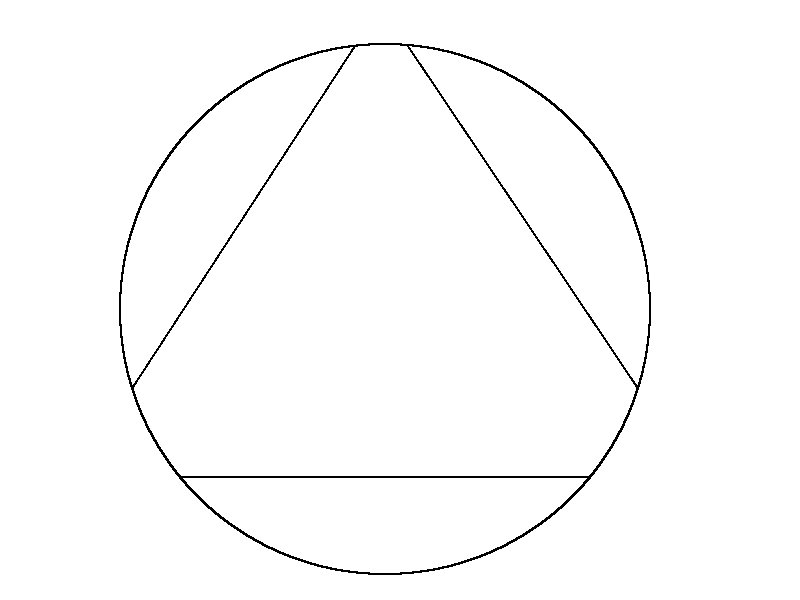}}}}
&
{\rotatebox{0}{\scalebox{0.15}{\includegraphics*{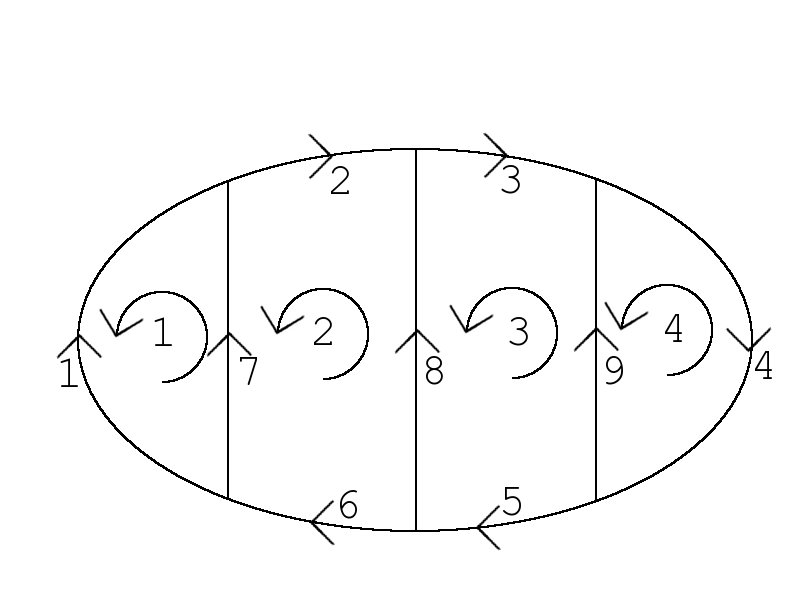}}}}
&
{\rotatebox{0}{\scalebox{0.15}{\includegraphics*{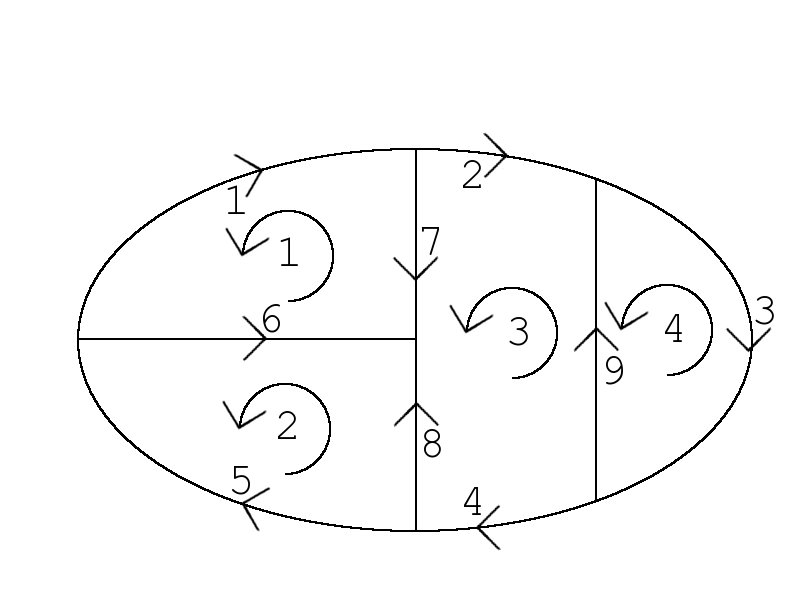}}}}
\\ (a)& (b)&(c)
\end{tabular}
\begin{tabular}{cccc}
&{\rotatebox{0}{\scalebox{0.15}{\includegraphics*{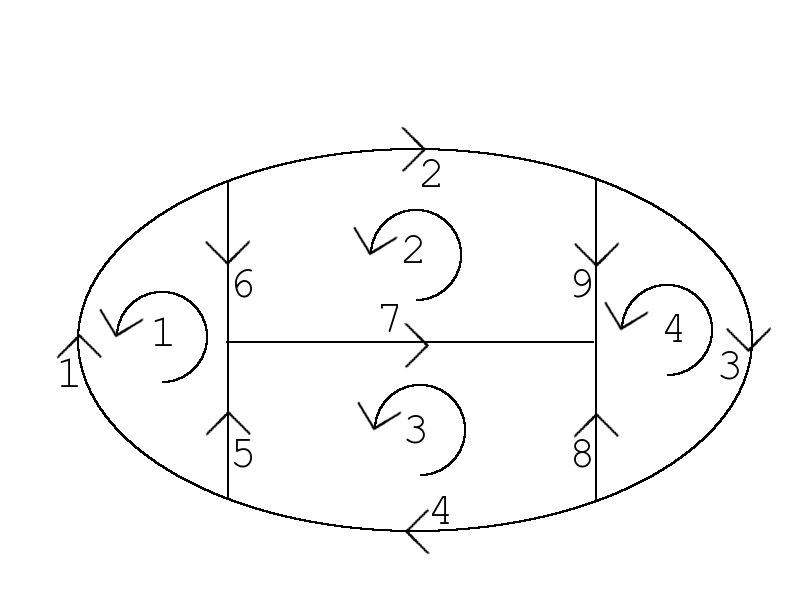}}}}
&
{\rotatebox{0}{\scalebox{0.15}{\includegraphics*{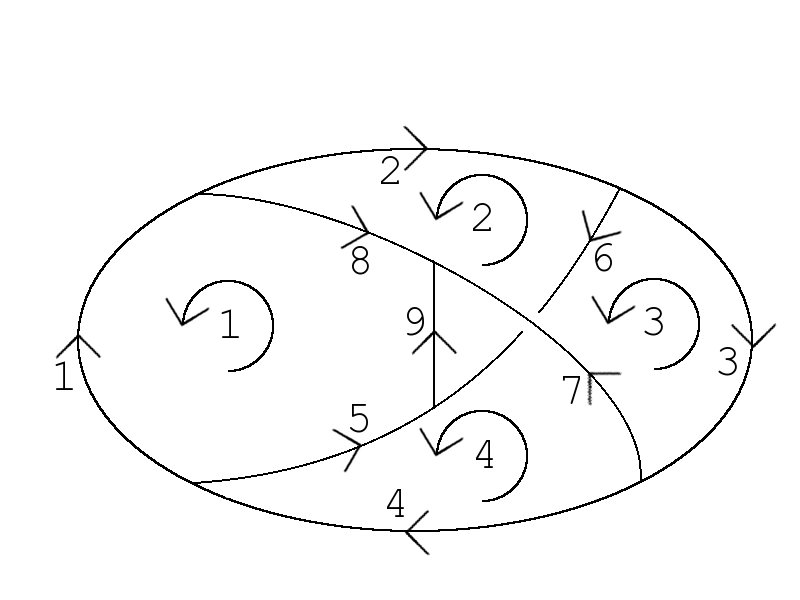}}}}&
\\ &(d) & (e)&\\
\end{tabular}
\caption{The five four-loop skeleton diagrams. }
\label{Figure:fourloop}
\end{center}
\end{figure}

The five distinct four-loop skeleton diagrams shown in figure~\ref{Figure:fourloop}  can be analysed in similar fashion.    In this case 20 $d$ zero modes must be supplied by the 9 $b$-ghost insertions together with the vertices. Once again,  in the supergravity case configurations with contact vertices contribute the same leading behaviour as those with single-particle vertices.  For brevity we will therefore restrict the following discussion to the insertion of single-particle vertices.

The diagram in figure~\ref{Figure:fourloop}(a) does not contribute to the four-point function because the vertex operators have to provide a total of six $d$ zero modes (two for each of the two-edge loops), which is not possible. 
 In the case of the ladder diagram of  figure~\ref{Figure:fourloop}(b)   the period matrix only has 7 independent entries (a general diagram  has 9 independent entries), so that the condition  \eqref{linearomega} is not satisfied.  This means that the 9 $b$ insertions in the  ladder diagram cannot all  contribute the maximum number of   $d$ zero modes via $b^{IJ}_H$ from  the second term in \eqref{bghostN1}.  This again follows from the Grassmann nature of the components of the zero mode tensor $b_H^{IJ}$, which prevents any two $b$ factors being proportional to each other.   As in the three-loop case, this restricts the assignment of $b$ zero modes, as follows.  One of the $b$ insertions on line 2 or line 6 in figure~\ref{Figure:fourloop}(b) has to contribute a single $d$ zero mode.  The same is true for the $b$ insertions on line 3 and line 5.  This generalises the discussion of the   three-loop case in an obvious manner.
   As a result, the $b$'s  can  contribute  at most $3,5,5$ and 3 $d$ zero modes on loops $1,2,3$ and 4, respectively, which is a total of 16 $d$'s. This means that  the single-particle vertices must contribute the remaining 4 $d$ zero modes (as well as 4 $\hat d$'s)  so that only  the $d\, W\, \hat d$ component of $V$ contributes in all four vertices.  Furthermore,  one pair of vertices must be attached to loop 1 and a second pair to loop 4.  Simple dimensional counting then determines the leading low momentum dependance of the ladder amplitude to be proportional to $\partial^{12}\cR^4\, \Lambda^{4D-26}$.   A similar analysis applied to the skeleton of figure~\ref{Figure:fourloop}(c) leads to the requirement that all four vertices must contribute $d$ (and $\hat d$) zero modes.  Two of these must be attached to loop 4 and  one each to loops 1 and 2.  The leading contribution to the amplitude  is proportional to $\partial^{10}\cR^4\, \Lambda^{4D-24}$. 

The skeletons in figures \ref{Figure:fourloop}(d) and  \ref{Figure:fourloop}(e)  have period matrices with ten independent entries so that  \eqref{linearomega} is satisfied, which  allows all  9 $b$ insertions to contribute 2 $d$ zero modes via $b^{IJ}_H$.  For figure~ \ref{Figure:fourloop}(d) this gives $4,5,5$ and 4  $d$ zero modes  on loops $1,2,3$ and 4, respectively, which is a total of 18 $d$'s. Therefore, two vertices do not need to contribute with the maximum number of $d$ zero modes (nor the maximum number of $\hat d$ zero modes) and can contribute four insertions of $P^m$ via the first term in \eqref{intvert}.  These become two powers of $P^2$ after the functional integration. This diagram contributes to the leading order at low energy.    This configuration gives 9 factors of $r$, and hence 9 covariant derivatives $D^9$ (with a factor of $\hat D^9$ coming from $\hat r^9$), although configurations with lower powers of $r$ also contribute to leading order at low energy\footnote{In principle a nonleading contribution with 11 $r$'s is possible, which is the total number needed,  but the coefficient of this term vanishes (as discussed in the context of the pure spinor string in \cite{Berkovits:2006vi}).}.  The only constraint on the positions of the  vertices is that one must be located on the first loop and another on the fourth loop due to the zero mode counting. A similar conclusion applies to the nonplanar contribution in figure~\ref{Figure:fourloop}(e).    The expression for the amplitude in these cases is proportional to the skeleton diagram with scalar vertices attached and four factors of loop momenta in appropriate places multiplied by the fermionic and pure spinor integral, which gives an expression of the form
\be
A^{(4)} \sim \left.\left<\lambda^3\hat{\lambda}^3D^9\hat{D}^9 E^2 \hat{E}^2\right>\right|_{\theta^5\hat{\theta}^{5}}\,  \, \Lambda^{4D-22} \sim
D^{16}\, \hat D^{16}\,G^4\, \Lambda^{4D-22}
\sim \sigma_2^2\, \cR^4\ \, \Lambda^{4D-22} \,.
\label{nonprotect}
\ee
This  way of expressing the result (which is again in accord with the results of \cite{Bern:2009kd}) demonstrates the significant fact that the expression is an integral over the whole of superspace, which is in accord with the intuition that $\partial^8 \,\cR^4$ is a D-term and will not be protected against higher-order perturbative contributions, unlike the preceding cases.  This will be discussed explicitly shortly.   The critical dimension is seen from \eqref{nonprotect} to be $D_c^{(4)} = 22/4$.

\begin{figure}[t!]
\begin{center}
{\rotatebox{0}{\scalebox{0.2}{\includegraphics*{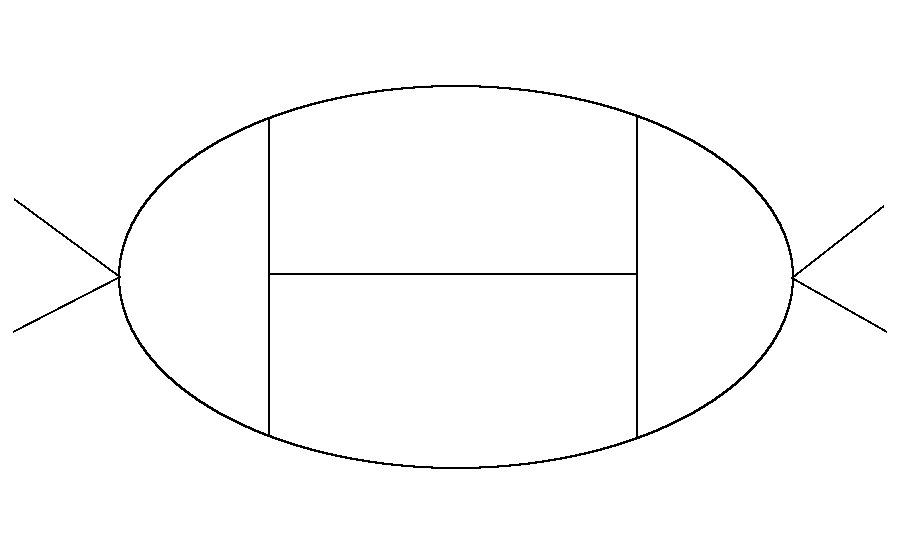}}}}
\caption{A four-loop diagram with two contact terms that gives the leading behaviour of $\partial^2\, \Tr F^4$.}
\label{Figure:fourcontact}
\end{center}
\end{figure}

In the Yang--Mills case the contact vertices are crucial in determining the leading behaviour of the single-trace amplitude in the large-$N$ (planar) limit.  This arises from the skeleton in figure~\ref{Figure:fourloop}(d), with  two contact vertices attached, each coupling two external particles to a point on the skeleton (see  figure~\ref{Figure:fourcontact}) as required by BRST invariance,  resulting in the behaviour $\partial^2\Tr F^4\, \Lambda^{4D-22}$.  Such contact vertices again do not contribute to the double-trace amplitude.  However,  in that case the presence of a pair of $P$'s leads to a $P^2$ insertion  and the double-trace amplitude is then proportional to $\partial^4(\Tr F^2)^2 \Lambda^{4D-24}$, as anticipated in \cite{Berkovits:2009aw}.  The nonplanar skeleton of  figure~\ref{Figure:fourloop}(e) also contributes the same leading behaviour as figure~\ref{Figure:fourloop}(d) but is suppressed by $1/N^2$ in the large-$N$ limit.

\medskip

{\bf (v)  Five loops}
\smallskip
\begin{figure}[h!]
\begin{center}
{\rotatebox{0}{\scalebox{0.2}{\includegraphics*{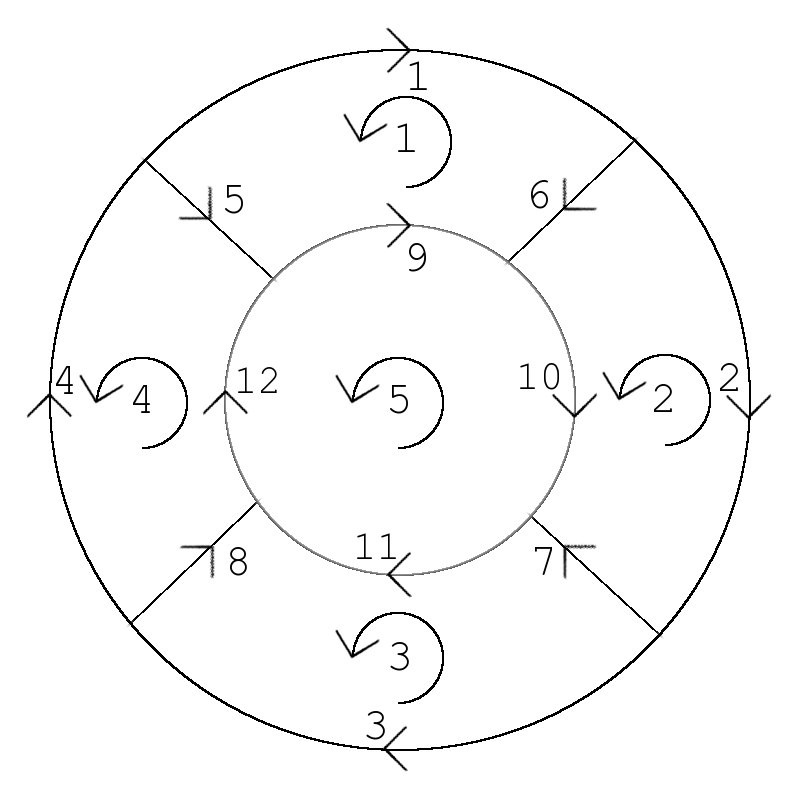}}}}
\caption{A five-loop skeleton for which there are at least twelve insertions of $r$.}
\label{Figure:five}
\end{center}
\end{figure}

\begin{figure}[t!]
\begin{center}
{\rotatebox{0}{\scalebox{0.8}{\includegraphics*{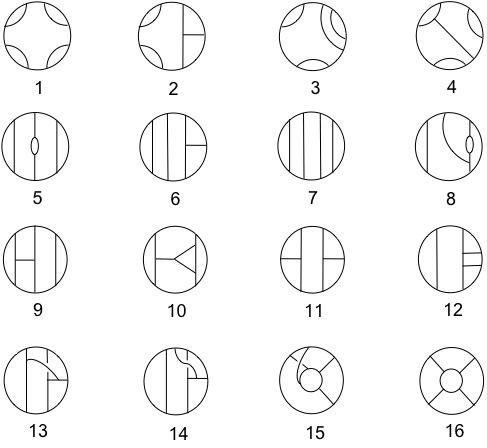}
}}}
\caption{The 16 independent  five-loop skeleton diagrams.}
\label{fiveloopskeleton}
\end{center}
\end{figure}

At $L=5$ there is a radical change in the pattern of the four-point function. In this case there are sixteen distinct skeleton diagrams, which are shown in figure~\ref{fiveloopskeleton}.   These are all possible vacuum diagrams in $\varphi^3$ field theory.      
We will focus on the diagram in figure~\ref{Figure:five} (which is number 16 in figure~7) for the present discussion.  It is notable that this diagram and diagram 15 in figure~7 are the only ones that have no loops with triangles or bubbles   As before, the four vertex operators need to be attached to this diagram, introducing four extra propagators,  and integrated over the whole skeleton. The number of $d$ zero modes that need to be provided by the four vertices and the twelve $b$-ghost insertions is 25,  or 5 for each loop.  The twelve $b$ insertions can provide  more than eleven $r$ insertions, which is more than the  required number of $r$ zero modes.   This happens, for example, if the second term in the $b$-ghost in \eqref{bghostN1} (the term quadratic in $d$) is used for all twelve $b$ insertions, which supplies twelve factors of $r\, d^2 /(\lambda\bar\lambda)$.  Figure~\ref{Figure:five} is an example of a skeleton for which terms with more than eleven powers of $r$ arise (this can be established using the structure of the period matrix of this diagram).  The nonzero $r$ modes can only contract with $s$'s  but there are none of these in the expression for the amplitude if the regulator, $\cN$, is ignored, in which case  the result apparently vanishes.  However,  there are also more than eleven powers of  $(\lambda\bar{\lambda})$ in the denominator, so the  $\lambda$, and $\bar \lambda$ integrals apparently diverge.   This means that there appears to be a new $0/0$ ambiguity, this time at small $\lambda$. The discussion of the $\hat{b}$ insertions is similar, resulting in the possibility of having twelve or more insertions of $\hat r$'s and at the same time twelve or more insertions of $(\hat{\lambda}\hat{\bar{\lambda}})$ in the denominator. The analogous five-loop phenomenon was noted in the pure spinor open string in \cite{Berkovits:2009aw}.

The regularisation of singularities of this type in pure spinor string theory was discussed in  \cite{Berkovits:2006vi,Aisaka:2009yp} where a complicated procedure involving the addition of extra fields was suggested. This approach should also be relevant for the pure spinor particle. The least complicated part of this procedure is the regularisation of the $r$ insertions, which can also be understood from the form of the large-$\lambda$ regulator $\cN$ defined in \eqref{regulator}.   The factors of $s$ that are needed to contract out the surplus $r$'s are contained in this regulator in the combination
\eqnb
S \, \lambda\, d \sim \lambda\, \bar \lambda\,  s\, d \,,
\label{regsmall}
\eqne
where $S_{mn} $ and $S$ are defined in \eqref{gaugecomb}.
Contraction with $r$ leads to a factor of $\lambda\, \bar \lambda\,  d$.  In other words,  the net result is to  trade an $r/\lambda \bar \lambda$ for a further factor of  $d$, thereby reducing the twelve powers of  $r/\lambda\bar\lambda$  to eleven.   This resolves the problem of having too many $r$'s but it is important to stress that the regulator $\cN$ does not regulate the resulting logarithmic divergence of the $\lambda$ and $\bar{\lambda}$ integrals. To do this, one needs the full structure of the complicated regulator discussed in \cite{Berkovits:2006vi,Aisaka:2009yp},  which still has the property that each $r/\lambda \bar \lambda$ is traded for a further factor of $d$. 
 
The specific example discussed above is one where each $b$ contributed a pair of $d$'s via the second term in 
\eqref{bghostN1} giving a total of 24 $d$'s and 12 $r$'s.  The final $d$ is provided by the regularisation procedure trading the extra $r$ for a $\lambda \bar \lambda d$, giving a total of $25$ $d$'s and 11 $r$'s.
With this configuration  all four vertex operators contribute via the term $P^m\, G_{mn}\, P^n$.   A similar conclusion is reached by inserting other terms in the $b$-ghost with fewer powers of $d$ but more powers of $r$.  For example, there are terms contributing $r^{13}\, d^{23}$,   $r^{14}\, d^{22}$, etc.  All of these reduce to $r^{11}\, d^{25}$ by repeatedly exchanging an $r$ for a $d$ until there are 11 powers of $r$.  In all cases the dominant behaviour at low energy is obtained by using the $P^m\, G_{mn}\, P^n$ part of the vertex.
 %%%%%%%%%%%%%%%%%%%%%%%%%%%%%%%%%%%%%%%%%%%%%%%%%%
\null
\vskip 0.5cm
 
  Upon performing the functional integral the eight insertions of $P^m$ produce a term proportional to four $P^2$ insertions on internal lines, as in \eqref{modP}.    For the skeleton shown in figure~\ref{Figure:five} the result is particularly simple.  By repeated use of the expression for contracting a pair of $P$'s, \eqref{contractp}, it is straightforward to see that the four $P^2$ effectively cancel propagators adjacent to each of the four vertex insertions.
 The resultant bosonic loop is simply the skeleton diagram itself,  whereas in the cases with $L<5$ the amplitude always had extra propagators compared to the skeleton.  The ultraviolet behaviour of the diagram in figure \ref{Figure:five}  for large enough  $D$ is simple to determine since there are 12 propagators and five $D$-dimensional loop integrals.  The degree of divergence is therefore $\Lambda^{5D-24}$, where $\Lambda$ is a high momentum cutoff.  
 
  This multiplies the fermionic and pure spinor integrals,  giving the leading low energy behaviour of the five-loop amplitude
\eqnb
A^{(5)} \sim \left.\left<\lambda^3\hat{\lambda}^3D^{11}\hat{D}^{11}G^4\right>\right|_{\theta^5\hat{\theta}^5} \, \Lambda^{5D-24}\sim D^{16}\, \hat D^{16}\, G^4\, \Lambda^{5D-24} \sim \sigma_2^2\, \cR^4\ \, \Lambda^{5D-24}
\,.
\label{fiveloopprefactor}
\eqne
Therefore, the term in (\ref{fiveloopprefactor}) has the same dimension as $\partial^{8}\cR^4$.  This signals a logarithmic ultraviolet divergence in $D=D_c^{(5)}=24/5$ dimensions, which is lower than the value $D=26/5$ expected if  $\partial^{8}\cR^4$ were protected against a five-loop contribution. This gives further evidence that $\partial^{8}\cR^4$ is a D-term and should get contributions from all loops.

 In addition to the contribution from the skeleton diagram of figure~\ref{Figure:five} there could be other skeleton diagrams that also contribute to the same leading behaviour of the amplitude.  We have not undertaken an exhaustive study of all sixteen skeleton diagrams, which are shown in figure~\ref{fiveloopskeleton} (the skeleton diagram of figure~\ref{Figure:five} being number 16).  However, there is at least one more skeleton that contributes to this leading behaviour, namely, number 15.  It is notable that skeletons 15 and 16 are the only ones that have no bubble or triangle subdiagrams.
 
 In the Yang--Mills case  the single-trace contributions come from diagrams with the planar subset of the skeletons shown  in figure 7 (i.e., excluding those numbered 13, 14, 15, which are nonplanar), whereas all the diagrams contribute to the nonplanar double-trace part of the amplitude.  The general pattern that follows from the above rules leads to the expected low energy behaviour $\partial^2 \Tr F^4\, \Lambda^{5D-26}$  for the single-trace term   (although our  statements are again based on only a partial analysis of the set of diagrams in figure~7).  As in the $L=3,4$ cases, the contact terms required by BRST invariance play an important r\^ole in obtaining this behaviour~\cite{jonas}.  They do not affect the double-trace term, which behaves as  $\partial^4(\Tr F^2)^2\, \Lambda^{5D-28}$  at low energy.  This is in agreement with the rather less detailed analysis based on the open string in \cite{Berkovits:2009aw} and with superspace arguments presented in that reference.  The systematics described here should provide insight into the way in which the many planar five-loop Yang--Mills diagrams listed in~\cite{Bern:2007ct} contribute to the leading ultraviolet behaviour of the amplitude.
 
One could ask whether the leading five-loop contribution from one skeleton might cancel with  contributions from other skeletons.  Although we cannot rule this out, this  would be very different from the situation at three and four loops \cite{Bern:2007hh,Bern:2009kd} since those  cancellations  are between terms in the same skeleton.  As we have seen in the $L\le 4$  cases, the result of any such cancellations  is manifest in the pure spinor formalism, which builds in the constraints of maximal supersymmetry, and each skeleton is treated independently.  An apparently accidental cancellation between different five-loop skeleton integrals would not be attributable to supersymmetry in any conventional sense, and would therefore be surprising.

\medskip
{\bf (vi) Higher loops, $L>5$}
\smallskip

We shall now  sketch an argument that  $\partial^{8}\cR^4$ gets contributions from all higher loops and is indeed unprotected.   For simplicity, assume that the relevant piece of the $b$-ghost is the second term in \eqref{bghostN1} that contains two $d$'s  (any of the terms in the $b$-ghost leads to the same conclusions as long as there are more than eleven insertions of $r$ and at least  $5L$ $d$'s, after transforming the surplus $r$'s into $d$'s). This gives an insertion of the form 
\be
\prod_{i=1}^{3(L-1)}\int_0^{T_i}\frac{d\tau}{T_{i}} \frac{1}{(\lambda\bar{\lambda})^2}\left(\bar{\lambda}\gamma_{mnp}r\right)\left(d\gamma_{mnp}d\right) \,,
\label{insertt}
\ee
 and thus involves $3(L-1)$ $r$'s.  Soaking up 11 zero modes leaves $3(L-1)-11$ $r$'s  that must be contracted with $s$'s in the small-$\lambda$ regulator,  producing a term $(\lambda\bar{\lambda}d)^{3(L-1)-11}$. Therefore, the $b$-ghost insertion is proportional to $d^{4(L-5)+5L}$.   But only $5L$ $d$'s are needed from the $b$-ghost insertions and the vertices, so  $4(L-5)$ $d$'s  can be contracted among themselves giving $2(L-5)$ insertions of $P^{m}$. The same argument holds for the $\hat{b}$-ghost, giving a total of $4(L-5)$ insertions of $P^{m}$.

Since the $b$ and $\hat{b}$ insertions contribute all missing $d$ and $\hat{d}$ zero modes ($5L$  of each), all four vertices can contribute via the term $P^mG_{mn}P^{n}$, giving another eight factors of $P_m$. This results in a  total of $4(L-3)$ insertions of $P^m$. Upon performing the functional integral these $4(L-3)$ $P^m$ insertions can produce a term with  $2(L-3)$ insertions of $P^2$ on internal lines. This gives the contribution at low energy and is proportional to the skeleton diagram with $2L- 10$ insertions of $P^2$. This multiplies the fermionic and pure spinor term giving the leading low energy behaviour
\eqnb
\hspace{-.3cm}
A^{(L)} &\sim& \left.\left<\lambda^3\hat{\lambda}^3D^{11}\hat{D}^{11}G^4\right>\right|_{\theta^5\hat{\theta}^5}\, \Lambda^{L(D-2) -14}
\sim D^{16}\, \hat D^{16}\, G^4 \, \Lambda^{L(D-2) -14} \no
&\sim& \partial^{8}\cR^4\, \Lambda^{L(D-2) -14}\,,
\eqne  
which is the same interaction as for the five-loop amplitude.  The logarithmic divergence in this case occurs in $D=D_c^{(L)}= 2 + 14/L$ dimensions, as anticipated in the introduction.  In particular, the divergence  occurs in $D=4$ dimensions at $L=7$ loops.

\section{Discussion}
\label{discussion}

In this paper we have analysed the ultraviolet properties of  four-particle loop amplitudes of maximally supersymmetric Yang--Mills and supergravity by use of a formulation of pure spinor quantum mechanics, the details of which are given  in \cite{jonas} and which is modelled on the pure spinor string.   We did not attempt to calculate the precise values of these amplitudes, but were concerned with their general structure.  The $L$-loop amplitude consists of a sum of terms associated with distinct $\varphi^3$ skeleton diagrams.  
 Integration of the  fermionic and pure spinor  zero modes  for a given skeleton leads to a factor of  of order
 $\partial^{2\beta_L}\, \cR^4$, which multiplies a multiloop scalar field theory diagram. This diagram is the skeleton to which four scalar vertices are attached  together with certain insertions of numerator momentum factors in a manner that is correlated with the zero mode integrations.  
 
 In this way we were able to reproduce the general structure of the diagrams that contribute to the four-graviton amplitude in explicit calculations up to four loops~~\cite{Green:1982sw,Bern:1998ug,Bern:2007hh,Bern:2009kd}.
 In particular, for a given $L$, we obtained the form of the leading ultraviolet divergence that arises in the critical dimension, $D_c^{(L)}$, and which is associated with the lowest order term in the low energy expansion of the amplitude.
 At five loops we found a change in behaviour associated with the fact that the number of $b$-ghost insertions required produces terms that possess more than eleven powers of $r/(\lambda \bar\lambda)$, such as those associated with the skeleton of figure~\ref{Figure:five}, that give the leading term in the low energy limit.  In order to complete the argument we needed to make mild use of the regulator for the pure spinor string introduced in \cite{Berkovits:2006vi} in order to regulate logarithmic divergences in the bosonic pure spinor integrals. This analysis pinpoints the terms responsible for the worst ultraviolet behaviour.
Since we have found that the leading ultraviolet dependence of the loop amplitudes based on the pure spinor particle  agrees with earlier calculations for   $L\le 4$, we are confident that some skeleton diagrams give nonvanishing contributions to $\partial^8\,\cR^4$ at $L\ge 5$ loops.
Nevertheless, as noted earlier, there is a possibility of cancellations between different skeletons.  

The considerations of this paper are almost all directly applicable to the pure spinor closed superstring.  In particular, the detailed counting of zero modes is almost identical, again resulting in contributions to the $\partial^8\cR^4$ interaction from tree-level and all loops.

The findings of this paper and other arguments based on string theory (summarised in \cite{Green:2010sp})  suggest that  $\beta_L=4$ for $L\ge 4$ (where $\beta_L$ was defined by \eqref{loopamp}), so that $\partial^8\, \cR^4$ is unprotected from perturbative contributions beyond four loops.  
If this is  case the first ultraviolet divergence can appear in $D=4$ maximal supergravity ($\cN =8$ supergravity) at seven loops, where $\partial^8\,\cR^4$ is a possible counterterm.   There are also arguments suggesting the possibility of a seven-loop divergence based on the study of supersymmetry constraints on counterterms in $\cN=8$ supergravity~\cite{Howe:1980th,Bossard:2009mn}.

\vspace{.5cm}
\noindent
{\bf{Acknowledgements}}
As this work was being finalised we received a draft of a paper  by Pierre Vanhove with related observations~\cite{Vanhove:2010aa}  based on the structure of loop amplitudes presented in~\cite{Bern:2010ue}.  We have also become aware of recent lectures by Nathan  Berkovits \cite{BerkovitsICTP} on closely  related topics.   JB acknowledges the support of the Swedish Research Council under project no.\ 623-2008-7048 and MBG is grateful for the support of European Research Council  Advanced Grant No. 247252.

%\vfill\eject

\end{document}